\documentclass[english]{article}
\usepackage{jcappub}
\usepackage{lmodern}

\usepackage{graphicx}
\usepackage{esint}
\usepackage{babel}
\usepackage{array}
\usepackage{longtable}
\usepackage{booktabs}
\usepackage{titlesec}
\titleformat{\section}
  {\normalfont\Large\bfseries}{\thesection}{1em}{}[{\titlerule[1.8pt]}]
\titleformat{\subsection}
  {\normalfont\Large\bfseries}{\thesubsection}{1em}{}[{\titlerule[0.8pt]}]
\makeatletter

\providecommand{\tabularnewline}{\\}

\makeatother

\usepackage{babel}
\begin{document}

\title{SI Toolbox - Full documentation}
\author[a,b]{Santanu Das}
\emailAdd{sdas33@wisc.edu}

\affiliation[a]{Department of Physics, University of Wisconsin - Madison, Madison, WI 53706, USA}
\affiliation[b]{Fermi National Accelerator Laboratory, Batavia, IL 60510, USA}
\version{Beta}

\keywords{}

\abstract{
SI Toolbox is a package for estimating the isotropy violation in the CMB sky. It can be used for estimating the BipoSH coefficients, Dipole modulation and Doppler boost parameters etc. 
Different Fortran subroutines, provided with this package, can help the users to develop their independent Fortran codes. This document is an overview of the SI Toolbox installation 
guide, stand-alone facilities and  Fortran subroutines.
}

\contribution{
The Package SIToolBox is coded up by Santanu Das. However, the work is based on multiple research papers~\cite{Das2015, Das2018, Pant2016, Das2016}. Here, I have listed everyone, who were directly or indirectly involved in developing the package SIToolBox. 

\begin{itemize}
\item{\bf Santanu Das} The research has been carried out and the software package SIToolBox has been developed by Santanu Das.
\item{\bf Shabbir Shaikh} He helped in finding a missing factor of $2$ in the original work~\cite{Das2015}. He also helped in multiple bug fixing and testing the algorithm by running it on multiple data-sets.  
\item{\bf Benjamin D. Wandelt} He initially proposed the algorithm for \texttt{bestmitor} and \texttt{betaestimator}, which has been discussed in detail in \cite{Das2015}.
\item{\bf Tarun Souradeep} He proposed the project~\cite{Das2015} and contributed in several interesting discussions related to the project. 
\item{\bf Nidhi Panth and Aditya Rotti} The subroutine \texttt{CalcBipoSH} is based on some of the software package, initially developed in cooperation with Nidhi Panth and Aditya Rotti \cite{Pant2016, Das2016}. They also helped the project by participating in multiple interesting discussions during the project.
\item{\bf Suvodip Mukharjee} Some of the nSI maps used for testing the algorithm are generated by the software package CoNIGS~\cite{Mukherjee2014} developed by Suvodip Mukharjee. He also helped the project by participating in several interesting discussions during the project.
\end{itemize} 
}
\maketitle

\section{Introduction}
Spherically distributed data with a random fields occur in many areas, including astrophysics, geophysics, optics, 
image processing and computer graphics. In many cases, it is the Gaussian random field, especially for geophysics 
and astrophysics. For a statistically isotropic Gaussian random field on a sphere, the two-point
correlation function is rotationally invariant and hence the co-variance
matrix of the corresponding random spherical harmonic coefficients,
i.e.  $\left\langle a_{lm}^{*}a_{l'm'}\right\rangle $ is diagonal
and independent of the azimuthal multipole index $m$. However, in
presence of SI violation, the co-variance matrix $\left\langle a_{lm}^{*}a_{l'm'}\right\rangle $
can depend on $m$ and the off-diagonal components can be nonzero.
Hence we require the BipoSH spectra ($\tilde{A}_{ll'}^{LM}$) to represent the complete statistics~\cite{Das2015}.
We develop the package SI Toolbox for analyzing the co-variance matrix of cosmic microwave
background (CMB) sky maps to infer the statistical isotropy (SI) of
our observed universe. However, the package is applicable to Bayesian
inference for any other studies involving a scalar random field on the sphere. 

SI toolbox is developed for calculating the full posterior distribution of angular power spectrum ($C_l$) and the BipoSH 
coefficients ($\tilde{A}_{ll'}^{LM}$) using Monte Carlo sampling without any marginalization over the spherical harmonic coefficients ($a_{lm}$).
The package provides multiple pre-compiled stand alone packages like \texttt{betaestimater}
(estimating the Doppler boost parameter, $\beta$,  assuming the isotropy violation in the sky
is completely due to the Doppler effect), \texttt{bestimator} (Bayesian estimator
of the BiopoSH coefficients from non-isotropic CMB sky map), \texttt{map2fits}, \texttt{fits2d}
(map conversion package from fits to ASCII format), \texttt{nest2ring}, \texttt{ring2nest}
(map conversion from nested format to ring format and vice versa), \texttt{rotateCoor} 
(For rotating the coordinate system from ecliptic to galactic etc.),
\texttt{clebshgen} (for calculating the Clebsch Gordan coefficients file). Apart
from these precompiled codes there are some Fortran functions for
calculating the BipoSH coefficients, Clebsch Gordan coefficients etc.
which can be called from different external programs.

\section{SI Toolbox Download Guideline}

SI Toolbox comprises a suite of Fortran 90 routines both stand-alone facilities and callable subroutines as an alternative for those users who wish to build their own tools. 
The distribution can be downloaded as a zipped file from 

\vspace{1 em} 
\noindent \url{https://github.com/SIToolBox/SIToolBox}\,.
\vspace{1 em} 

\noindent It will give you a zipped file named \texttt{SIToolBox-master.zip}, which can respectively be unpacked and renamed by executing the commands 

\begin{verbatim}
% unzip SIToolbox-master.zip
% mv SIToolBox-master/ SIToolBox 
\end{verbatim}

\begin{figure}[h]
\centering
\includegraphics[width=0.9\textwidth]{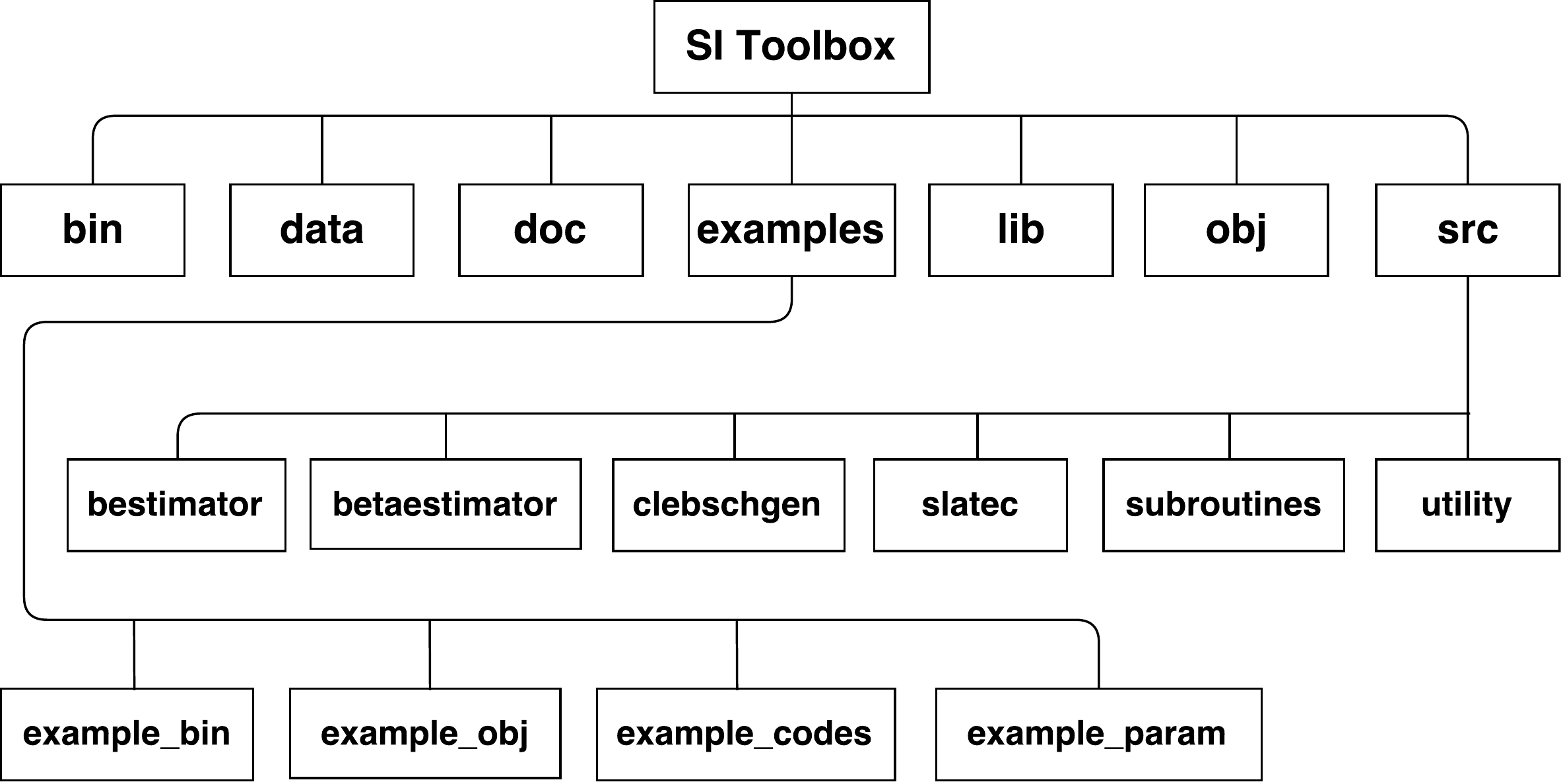}

\caption{\label{fig:DirectoryStructure}The directory structure for SIToolbox distribution. }
\end{figure}

\noindent This will create a directory named SIToolbox. The directory structure is shown in Fig.~\ref{fig:DirectoryStructure}.

\newpage
\section{SIToolbox directory structure}
The directory structure of SI Toolbox is shown in Fig.~\ref{fig:DirectoryStructure}. Here we broadly discuss the directory structure. 

\begin{itemize}
\item \textbf{\texttt{SIToolbox/bin}}: It is a standard sub-directory that contains the executable (i.e., ready to run) programs. Initially this directory will be empty. After running the \texttt{./compile.sh} command it will store the executable. 
\vspace{1 em} 
\item \textbf{\texttt{SIToolbox/obj}}: It stores temporary object files while compiling the code. The object files can be removed once compilation is done. 
\vspace{1 em} 
\item \textbf{\texttt{SIToolbox/lib}}:  This subdirectory contains the SIToolbox library files. Initially it will be an empty directory. After running \texttt{./compile.sh} it will store two library files namely \texttt{libslatec.a} and \texttt{libsubroutines.a}. Users, who like to write their own packages using the callable SIToolBox subroutines need to include these libraries while compiling their codes. 
\vspace{1 em} 
\item \textbf{\texttt{SIToolbox/doc}}: This subdirectory contains the documentation of the SI Toolbox package and the related papers. 
\vspace{1 em} 
\item \textbf{\texttt{SIToolbox/data}}: This subdirectory suppose to contain some non-isotropic maps and Clebsch Gordan coefficient files, bias function for the Doppler modulation and the $C_l$ corresponding to the bias function etc. These can be used for initial testing the proper compilation before running it on some real dataset. Presently no dataset is there in this directory due to file size restriction in github. The Clebsch Gordan coefficients can easily be generated using the \texttt{clebschgen} routine provided with this package. 
\vspace{1 em} 
\item \textbf{\texttt{SIToolbox/source}}: This sub-directory contains the source codes. There are 6 sub-directories inside this folder, namely \texttt{bestimator},  \texttt{betaestimator}, \texttt{clebschgen}, \texttt{utility}, \texttt{slatec}, \texttt{subroutines}. The details of the first 4 are discussed in Sec.~\ref{standalone} and the last two are discussed in Sec.~\ref{callable}.
\vspace{1 em} 
\item \textbf{\texttt{SIToolbox/examples}}: Inside this directory there are few example codes which can help those who wants to develop their own packages. 
\end{itemize}

\newpage
\section{SI Toolbox Installation Guideline}

SI toolbox is written in \texttt{Fortran 90}. Therefore, you need some fortran compiler and \texttt{openmp} library for compiling the program .
The package also uses \texttt{HEALPix}~\cite{Gorski2004} and \texttt{CFITSIO} library. 

\vspace{1 em} 
\subsubsection*{Required packages}
\vspace{1 em} 
\begin{itemize}
\item \textbf{\texttt{HEALPix}}: \texttt{HEALPix} is a map projection software and it is extensively used for CMB data analysis. It
can be downloaded from  
\url{http://healpix.sourceforge.net/}

You need the library files \texttt{libhealpix.a}, \texttt{libhpxgif.a}.
\vspace{1 em} 
\item \textbf{\texttt{CFITSIO}}: \texttt{CFITSIO} is a library of C and Fortran subroutines for reading and writing data 
files in FITS (Flexible Image Transport System) data format. It
can be downloaded from  
\url{https://heasarc.gsfc.nasa.gov/fitsio/fitsio.html}

You need the library files \texttt{libcfitsio.a}.
\vspace{1 em} 
\item\textbf{\texttt{OpenMP}}: The parallelization  is done with \texttt{OpenMP}. You need the files \texttt{fopenmp} and \texttt{xopenmp} for compiling SIToolBox.

\end{itemize}

\vspace{1 em} 
\noindent All these packages should be available in your system before compiling SIToolBox. 
 
For compiling the package, go to the SIToolBox folder. Open the \texttt{compile.sh} file in some text editor and set the path of your local \texttt{F90} compiler, \texttt{HEALPix}, 
\texttt{CFITSIO}, \texttt{OpenMP} library files and \texttt{HEALPix} include files. It should look something as follows

\vspace{1 em} 
\begin{verbatim}
# F90 Compiler
FC="/../intel/mpich-3.0.4/bin/mpif90"

#INCLUDE Path, LIB Path, FLAG Path
INCLUDE="-I/../Healpix_2.15a/include"
LIB="-L/../Healpix_2.15a/lib -L/../cfitsio -lhealpix -lhpxgif -lcfitsio"
FLAG="-fopenmp"
-------------
\end{verbatim}
\vspace{1 em} 

\noindent Give execute permission to the file \texttt{compile.sh} using the following command
\vspace{1 em} 
\begin{verbatim}
chmod a+x compile.sh
\end{verbatim}
\vspace{1 em} 
Once \texttt{compile.sh} has the execute permission just run \texttt{compile.sh} using the command
\vspace{1 em} 
\begin{verbatim}
./compile.sh
\end{verbatim}
\vspace{1 em} 
\noindent This will compile all the main codes. There are few example codes inside the folder \texttt{/examples} 
For compiling the example codes first give the execute permission to the \texttt{compile.sh} file inside \texttt{/examples} directory, change the 
\texttt{F90} compiler, \texttt{HEALPix}, \texttt{CFITSIO}, \texttt{OpenMP} library path in \texttt{compile.sh}
and then run the command \texttt{./compile.sh} inside that folder. It will compile all the example codes.

\newpage
\section{\label{standalone}Available standalone processes}

\subsection{betaestimator}

This facility privides a means to generate the Monte Carlo chains
for the Doppler $\beta$ parameter from a non-isotropic skymap assuming that the isotropy violation
in the skymap is only due to the Doppler boost \cite{Das2015, Das2018}. However, this facility
can also be used for the estimation of the Dipole modulation parameter, 
where the equations are similar to the Doppler boost. 

\titulo{Location inside SIToolbox directory}

\noindent\texttt{SIToolbox/src/betaestimator/}

\titulo{Details of the codes} 
There are 4 FORTRAN programs inside the folder independent of each other (except the common subroutines they are calling). 
A brief detail of each of the program is given here so that users can customize the programs accordingly.  
\begin{itemize}
\item \noindent\textbf{\texttt{betaestimator.f90 :}} 
It is the wrapper program for reading the input parameter file and calling the appropriate subroutine.

\item \noindent\textbf{\texttt{beta\_estimation\_nonoise.f90 :}}
It generates the Monte-Carlo
chains for beta values when there is no noise in the map. In such
cases, we don't need to evaluate the $a_{lm}$s, as they
are fixed. So the process is very fast. In single processor it takes
just few minutes to finish the calculations. 

\item \noindent\textbf{\texttt{beta\_estimation\_isotropicnoise.f90 :}}
It generates the Monte-Carlo chains when there is some isotropic noise in the map. In such cases
we can calculate the noise variance inverse in the spherical harmonic
space making it less time consuming than the anisotropic case. 

\item \noindent\textbf{\texttt{beta\_estimation\_anisotropicnoise.f90 :}}
Generates the Monte-Carlo chains for beta parameter when noise in the sky is anisotropic or there
is masking or both. In such case it is not possible to invert the noise matrix in
the spherical harmonic space. So we need to go to the pixel
space and invert the matrix pixel by pixel. This process is time consuming
and runs about 3 times slower than \texttt{beta\_estimation\_isotropicnoise.f90}. 
\end{itemize}

\titulo{How to run}

\noindent{\texttt{\% betaestimator < betaestimator.in}}

\titulo{Example parameter file}

\noindent{\texttt{SIToolbox/examples/example\_param/betaestimator.in }}

\vspace{3ex}
\noindent Here we briefly discuss the details of the variables in the parameter file.

\begin{longtable}{>{\centering}m{0.25\columnwidth}>{\centering}m{0.25\columnwidth}>{\centering}m{0.42\columnwidth}}
\toprule 
Name & Example Value & Description\tabularnewline
\midrule
\endhead
\texttt{NOISE} &  \texttt{no-noise} & Type of Noise : (\texttt{no-noise} /  \texttt{isotropic} /  \texttt{anisotropic})\tabularnewline\\
\texttt{MASK} &  \texttt{yes} & Do you want masking?  (\texttt{yes} /  \texttt{no})\tabularnewline\\
\texttt{CLEBSCH\_PATH} & \path{/home/sdas33/DATA/Final_beta_estimation/clebsch/clebs.dat}  & Precalculated Clebsch Gordan Coefficients ($C_{l1l2m1m2}^{LM}$) file. \tabularnewline
\texttt{CLEBSCH\_Lmax} & \texttt{2}  & $L^{max}$ for the Clebsch Gordan  file (for which the file has been generated).\tabularnewline\\
\texttt{CLEBSCH\_l1max} &  \texttt{1024} & $l_{1}^{max}$ for the Clebsch Gordan file (for which the file has been generated). \tabularnewline\\
\texttt{SHAPE\_FACTOR\_PATH} &  \path{/home/sdas33/DATA/SIToolBox/examples/fs.d} & Shape factor file with the full path (shape factor must be written in Hazian-Souradeep format \cite{Hajian2003,
Joshi2012,Das2016})\tabularnewline\\
\texttt{MASK\_PATH} & \path{/home/sdas33/DATA/Data_SIToolBox/mask/wmapmask_E_con.d} & Mask map file name along with full path.
For no masking you can leave it blank. \tabularnewline\\
\texttt{CL\_PATH} & \path{Planck2015TTlowP_totCls.dat} & $C_l$ of the input map. In {\texttt betaestimator} we are not varying $C_l$. 
If you don't have the best fit $C_l$, then you can calculate it using \texttt{bestimator} and then use it for betaestimator. \tabularnewline\\
\texttt{PIXEL\_WINDOW\_FUNCTION} & \path{pixel_window_n0512_t1.txt} & HEALPix pixel window function for the given $N_{side}$. \tabularnewline\\
\texttt{NOISE\_SD\_PATH} & \path{/home/sdas33/DATA/SIToolBox/data/Nmap.d} & Anisotropic Noise standard deviation map file name along with full path.
For isotropic noise you can leave it blank. \tabularnewline\\
\texttt{NOISE\_SD} &  \texttt{30.0} & Noise standard deviation in pixel space if \texttt{NOISE=isotropic}.
Otherwise you can keep it blank or set any random value. Unit is same as the unit used in the map.\tabularnewline\\
\texttt{MAP\_PATH} & \path{/home/sdas33/DATA/SIToolBox/data/Tmap.d}  & Enter the input map path. Map should be in ASCII format and ordering
is RING. (For fits files you need to convert it to ASCII format using \texttt{fits2d} command)\tabularnewline\\
\texttt{MAP\_NSIDE} &  \texttt{512} & \texttt{Nside} of the map\tabularnewline\\
\texttt{CHAIN\_PATH} &  \path{/home/sdas33/DATA/SIToolBox/examples/Bestimator_Chain/} & Location of the folder where the chain will be stored. One run can
generate only one chain. To generate multiple chains you can submit
the code multiple times but the Chain folder should be different for
each of the submission. Otherwise it will mix the values. \tabularnewline\\
\texttt{CHAIN\_Lmax}  & \texttt{2} & $L^{max}$ for the BipoSH chains. \tabularnewline\\
\texttt{CHAIN\_l1max}  & \texttt{1024} & $l_{1}^{max}$ for the BipoSH chains.\tabularnewline\\
\texttt{SAMPLE\_NUMBER}  &  \texttt{5000}  & Number of sample points for Monte-Carlo chain of the $\beta$ parameter. \tabularnewline\\
\bottomrule
\end{longtable}

\newpage
\subsection{bestimator}

This facility provides a means to generate the Monte-Carlo chains
for BipoSH coefficients from a non-SI skymap \cite{Das2015, Das2018}. 
We do not provide any facility to calculate the BipoSH chains for no noise 
case because there are trivial analytic solutions for that case. However, 
someone can easily simulate such cases by setting the noise variance 
to a negligibly small value. The values in the 
output file are written in Hazian-Souradeep format. 

\titulo{Location inside SIToolbox directory } 

\noindent\texttt{SIToolbox/src/bestimator/}

\titulo{Details of the codes} 
There are three Fortran programs inside the directory independent of the programs in other directory. 
So users can easily modify the programs according to their own requirements. Brief details of the programs are as follows.   
\begin{itemize}
\item \textbf{\texttt{bestimator.f90 :}} Its the wrapper for reading the input file and running the appropriate subroutine. 

\item \textbf{\texttt{BipoSH\_ALMll\_isotropic\_Noise : }}Generates the BipoSH Chains from a map with
isotropic noise. For isotropic noise, $N_{lml'm'}$ being a diagonal matrix, is invertible in the spherical harmonic
space making the process less time consuming.

\item \textbf{\texttt{BipoSH\_anisotropic\_noise :} }Generates the BipoSH chains when noise field
is anisotropic. Here, $N_{lml'm'}$ is not a diagonal matrix and hence the inversion is not 
possible in spherical harmonic space. So we go to the pixel space and invert the matrix pixel by pixel.
This process is time consuming and runs about 3 times slower then the isotropic noise case. 
\end{itemize}

\titulo{How to run}

\noindent{\texttt{\% bestimator < bestimator.in}}

\titulo{Input parameter file}

\noindent{\texttt{SIToolbox/examples/example\_param/bestimator.in }}

\vspace{3ex}
\noindent Here we briefly discuss the details of the variables in the parameter file.

\begin{longtable}{>{\centering}m{0.24\columnwidth}>{\centering}m{0.25\columnwidth}>{\centering}m{0.43\columnwidth}}
\toprule 
Name & Example Value & Description\tabularnewline
\midrule
\endhead
\texttt{NOISE} & \texttt{isotropic} & Type of Noise: (\texttt{isotropic} / \texttt{anisotropic}) \tabularnewline \\
\texttt{MASK} &  \texttt{yes} & Do you want masking?  (\texttt{yes} /  \texttt{no})\tabularnewline\\
\texttt{CLEBSCH\_PATH} & \path{/home/sdas33/DATA/Final_beta_estimation/clebsch/clebs.dat}  & Precalculated Clebsch Gordan Coefficients $C_{l1l2m1m2}^{LM}$ file with full path.\tabularnewline\\
\texttt{CLEBSCH\_Lmax} & {\texttt2} & $L^{max}$ for the Clebsch Gordan file.\tabularnewline\\
\texttt{CLEBSCH\_l1max} & \texttt{1024} & $l_{1}^{max}$ for the Clebsch Gordan file.\tabularnewline\\
\texttt{NOISE\_SD\_PATH} & \path{/home/sdas33/DATA/SIToolBox/data/Nmap.d} & Filename with the full path of the anisotropic noise standard deviation file if  \texttt{NOISE=anisotropic}. Otherwise leave it blank or put some arbitrary value.  \tabularnewline\\
\texttt{NOISE\_SD} & \texttt{30.0} & Noise standard deviation in pixel space, if \texttt{NOISE=isotropic}\tabularnewline\\
\texttt{MASK\_PATH} & \path{/home/sdas33/DATA/Data_SIToolBox/mask/wmapmask_E_con.d} & Mask map file name along with full path.
For no masking you can leave it blank. \tabularnewline\\
\texttt{MAP\_PATH} &  \path{/home/sdas33/DATA/SIToolBox/data/Tmap.d}  & Map file name with full path. Map should be in ASCII format and
ordering is RING.\tabularnewline\\
\texttt{PIXEL\_WINDOW\_FUNCTION} & \path{pixel_window_n0512_t1.txt} & HEALPix pixel window function for the given $N_{side}$. \tabularnewline\\
\texttt{MAP\_NSIDE} & \texttt{512} & HEALPix Nside of the map.\tabularnewline\\
\texttt{CHAIN\_PATH} & \path{/home/sdas33/DATA/SIToolBox/examples/Bestimator_Chain/anisotropic} & Location of the folder where the chain will be stored. One run can
generate only one chain. For multiple runs you must specify different chain folders. \tabularnewline\\
\texttt{CHAIN\_Lmax}  & \texttt{2} & $L^{max}$ for the BipoSH chains. \tabularnewline\\
\texttt{CHAIN\_l1max}  & \texttt{1024} & $l_{1}^{max}$ for the BipoSH chains.\tabularnewline\\
\texttt{SAMPLE\_NUMBER} & \texttt{5000} & Number of sample points in the BipoSH chains. You can also run multiple
small chains separately and then merge the files. \tabularnewline\\
\bottomrule
\end{longtable}

\titulo{Output chain files}
\noindent We follow the following conversion for the output chain file names 

\vspace{2ex}
\noindent \texttt{A(R/I)\_(LM)\_ll(d).d}
\vspace{2ex}

\noindent where, `\texttt{R}' or `\texttt{I}' stands for the real and the imaginary parts of the
coefficients. \texttt{L} and \texttt{M} are the \texttt{L} and \texttt{M} values of $A_{ll-d}^{LM}$
and \texttt{d} is the difference between $l_{1}$and $l_{2}$. So the file
\texttt{AR\_10\_ll1.d} stores real part of $A_{ll-1}^{10}$.

\newpage

\subsection{map2fits, fits2d, nest2ring, ring2nest, rotateCoor}
These are the utility facilities and provide handy means for pre/post processing of the data. These stand alone facilities are based on different HEALPix  subroutines.

The first two facilities, namely \texttt{map2fits} and \texttt{fits2d} provide a means to convert a map from ASCII to fits
and vice-versa. These conversions will preserve the ordering (\texttt{NESTED} or \texttt{RING}) of the input map. Except these two facilities 
all the other standalone facilities provided in \texttt{SIToolbox} run on the ASCII files. 

The next two facilities, namely \texttt{nest2ring} and \texttt{ring2nest} provide a 
means to change the ordering of the input map, from \texttt{RING} ordering
to \texttt{NESTED} ordering or vice-versa. All the estimator codes in \texttt{SIToolbox} run on the
\texttt{RING} ordering. So in many cases, converting the map between two ordering is important. 

The last facility, i.e. \texttt{rotateCoor} will convert the map from one coordinate system to another coordinate system. The input and output coordinate systems are \texttt{Galactic}, \texttt{Ecliptic}, \texttt{Celestial} and \texttt{Equatorial}.

\titulo{Location inside SIToolbox directory} 
\vspace{-3 ex}
\begin{longtable}{>{}m{0.45\columnwidth}>{}m{0.45\columnwidth}}
\endhead
\texttt{SIToolkit/src/utility/map2fits.f90} & \texttt{SIToolkit/src/utility/fits2d.f90}\tabularnewline
\texttt{SIToolbox/src/utility/nest2ring.f90}& \texttt{SIToolbox/src/utility/ring2nest.f90}\tabularnewline
\texttt{SIToolbox/src/utility/rotateCoor.f90} & \tabularnewline
\end{longtable}

\titulo{How to run}
You can run these facilities interactively using Linus shell. Here we show an interactive run of \texttt{fits2d} facility. Runs are in general self-explanatory.
\begin{verbatim}
% fits2d 
  Enter Nside for the map
  512
  Enter the input map
  map.fits
  Successful conversion
  Enter the output file name
  map.d   
\end{verbatim}
\titulo{Input parameters}

\begin{longtable}{>{\centering}m{0.35\columnwidth}>{\centering}m{0.63\columnwidth}}
\toprule 
Name & Description\tabularnewline
\midrule
\endhead
\texttt{Nside} & HEALPix Nside for the Input and Output map\tabularnewline
\texttt{Input Map} & File name with full path of the Input map. Except \texttt{fits2d} all the other inputs must be in ASCII format.\tabularnewline
\texttt{Output Map} & Output file name with full path. Except \texttt{map2fits} the output file should be in ASCII format everywhere else.\tabularnewline
\texttt{Ordering} & For \texttt{map2fits} you need to specify the ordering of the file for writing in the header. Choices are ( \texttt{1} / \texttt{2} ), where \texttt{1.RING} and \texttt{2.NESTED}. Default ordering is \texttt{RING} ordering. \tabularnewline
\texttt{Input/Output Coordinate} & For \texttt{rotateCoor} you have to specify the \texttt{input}/\texttt{output} coordinate system of the file. Options are \texttt{G. Galactic}, \texttt{E. Ecliptic}, \texttt{C. Celestial} and \texttt{E. Equatorial}. \tabularnewline
\bottomrule
\end{longtable}

\newpage
\subsection{clebschgen}

This facility privides a means to generate the Clebsch-Gordan coefficients
($C_{l_{1}m_{1}l_{2}m_{2}}^{LM}$) file for BipoSH calculation, \texttt{bestimator} and \texttt{betaestimator}. This facility uses the \texttt{slatec}\footnote{\url{http://www.netlib.org/slatec/}} library for calculating the
Clebsch-Gordan coefficients. It calculates all the nonzero Clebsch
Gordan coefficients, given a maximum value of $L^{max}$ and
$l_{1}^{max}$ and store them in a file. 

Note that, as we are writing all the values of Clebsch-Gordan coefficients in a file, it is necessary to read the full file before calculating a particular Clebsch Gordan coefficient. However, if you are interested in a particular Clebsch Gordan coefficient or a couple of them then you can directly call the clebsch stand alone function. 

\titulo{Location inside SIToolbox directory}

\noindent\texttt{SIToolbox/src/clebschgen/clebschgen.f90}

\titulo{How to run}
Here we present a simple run of the program \texttt{clebschgen}. Interactive run is self explanatory. We just need to provide $L^{max}$ and $l_1^{max}$ 
for the Clebsch-Gordan coefficients. The output file name will be chosen by the program. 
\begin{verbatim}
% clebschgen 
  Program : clebschgen
  It will calculate Clebsch Gordan coefficients C^{L M}_{l1 m1 l2 m2}
  Please input L_max and l1_max
  2 1024
  Output Clebsch filename : Clebs_Lmax_2_lmax_01024.dat 
\end{verbatim}

\titulo{Input parameters}
\begin{longtable}{>{\centering}m{0.3\columnwidth}>{\centering}m{0.64\columnwidth}}
\toprule 
Name & Description\tabularnewline
\midrule
\endhead
$\texttt{L\_max}$ & Maximum value of $L$ in $C_{l_{1}m_{1}l_{2}m_{2}}^{LM}$\tabularnewline\tabularnewline
$\texttt{l1\_max}$ & Maximum value of $l_{1}$ in $C_{l_{1}m_{1}l_{2}m_{2}}^{LM}$\tabularnewline
\bottomrule
\end{longtable}

\titulo{Output File}
\noindent \textbf{\texttt{Clebs\_Lmax\_{*}\_lmax\_{*}.dat}} :   It stores the $C_{l_{1}m_{1}l_{2}m_{2}}^{LM}$ values
up to $L^{max}$ and $l_{1}^{max}$. The file has two columns. The first column is the one dimensional index of the of the Clebsh Gordan coefficient
and the second column is the  Clebsh-Gordan coefficient. 
The one dimensional index of $C_{l_{1}m_{1}l_{2}m_{2}}^{LM}$
in the file is given by \texttt{ClbIndex} that can be obtained by calling
the subroutine 
  
\vspace{1 em} 
\noindent \texttt{Clebsch2OneD(L,M,l1,l2,m1,lmax,ClbIndex)}.
 
\vspace{1 em}   
\noindent We can open the file in \texttt{FORTRAN} by calling 

\vspace{1 em}   
\noindent \texttt{open(1,file='{*}{*}{*}{*}', action='read',status=OLD)}

\vspace{1 em}   
\noindent and read the file in an array by calling
\vspace{1 em} 
\begin{verbatim}
      read(1,*)recno,cleb
      Clebs(recno)=cleb
\end{verbatim}

\noindent We can get the \texttt{ClbIndex}-th value 
by  \texttt{Clebs(ClbIndex)}. Note that, the file only stores the values for $l_1 < l_2$. 
For $l_1 > l_2$ We need to use the properties of the Clebsch-Gordan coefficients to calculate it from $l_1 < l_2$ values.


\titulo{Example code for reading output file}

\vspace{-1 em}   
\noindent \texttt{SIToolBox/examples/example\_codes/test\_readClebs.f90}

\newpage
\section{Available callable subroutines}
\label{callable}
\subsection{lm2n, n2lm}

The first subroutine, i.e. \texttt{lm2n( )} is useful for storing two dimensional $a_{lm}$ into
a single dimensional array $q_{n}$. This is useful because if we
allocate $a_{lm}$ as
\begin{verbatim}
allocate(alm(0:lmax, 0:lmax))
\end{verbatim}
\selectlanguage{english}%
then half of the allocated spaces will not be used. However, if we write
it as a single dimensional array then it will help while passing the
variable into different functions or different MPI processors. Also, for calling the function \texttt{CalcBipoSH} 
we need one dimensional array of $a_{lm}$ that can be generated using \texttt{lm2n( )} subroutine. 

We can convert the one dimensional array back to two dimensional $a_{lm}$ array using \texttt{n2lm( )}.

\titulo{Format}

\noindent \texttt{call lm2n(l,m,n)                        }  \hspace{3 em} OR \hspace{3 em} \texttt{                        call n2lm(n,l,m)}

\titulo{Arguments}

\begin{longtable}{>{\centering}m{0.16\columnwidth}>{\centering}m{0.16\columnwidth}>{\centering}m{0.16\columnwidth}>{\centering}m{0.4\columnwidth}}
\toprule 
Name & Kind & In/Out & Description\tabularnewline
\midrule
\endhead
\texttt{l} & INT & IN / OUT & $l$ value of $a_{lm}$\tabularnewline\\
\texttt{m} & INT & IN / OUT & $\texttt{abs}(\texttt{m})$ value of $a_{lm}$. This variable can only take values from
$0$ to $l$.\tabularnewline\\
\texttt{n} & INT & OUT / IN & Array index in the single dimensional array. \tabularnewline\\
\bottomrule
\end{longtable}

\titulo{Example}
\vspace{-3 em}
\begin{longtable}{>{}m{0.45\columnwidth}>{\centering}m{0.15\columnwidth}>{}m{0.45\columnwidth}}
\begin{verbatim}
integer :: n,l,m
l=20 m=14     ! m must be positive 
call lm2n(l,m,n)
write(*,*) n
end program 
\end{verbatim}
& OR
&
\begin{verbatim}
integer :: n,l,m
n=224 
call n2lm(n,l,m)
write(*,*) l,m
end program 
\end{verbatim}
\end{longtable}

\titulo{Location of Example code}

\noindent\texttt{SIToolbox/examples/example\_codes/test\_lm2n.f90}

\noindent\texttt{SIToolbox/examples/example\_codes/test\_n2lm.f90}

\pagebreak{}

\subsection{Clebsch2OneD}

Saving the Clebsch-Gordan coefficients in one dimensional format is
useful instead of 6 dimensional matrices because most of the values
of the Clebsch-Gordan coefficients are zero. So instead of saving it
in 6 dimensional matrix if we store it in a one dimensional matrix
then it will save a lots of memory. It is also helpful while storing 
the array in a direct access file or reading it from there. 

\titulo{Format}
\begin{verbatim}
call Clebsch2OneD(L,M,l1,l2,m1,lmax,ClbIndex)
\end{verbatim}

\titulo{Arguments}

\begin{longtable}{>{\centering}m{0.16\columnwidth}>{\centering}m{0.16\columnwidth}>{\centering}m{0.16\columnwidth}>{\centering}m{0.4\columnwidth}}
\toprule 
Name & Kind & In/Out & Description\tabularnewline
\midrule
\endhead 

\texttt{L}, \texttt{M}
 & INT & IN & Array index of one dimensional representation of $C^{LM}_{l_1 m_1 l_2 m_2}$.\tabularnewline\\
\texttt{l1}, \texttt{l2}
 & INT & IN & $l_1$ and $l_2$ index of $C^{LM}_{l_1 m_1 l_2 m_2}$. \tabularnewline\\
\texttt{m1}
 & INT & IN & $m_1$ index of $C^{LM}_{l_1 m_1 l_2 m_2}$ \tabularnewline\\
\texttt{lmax}
 & INT & IN & Maximum value that $l_{1}$ can take\tabularnewline\\
\texttt{ClbIndex} & INT & OUT & Output one dimensional index of the Clebsch-Gordan coefficient. \tabularnewline
\bottomrule
\end{longtable}

\titulo{Example}
\begin{verbatim}
integer :: L,M,l1,l2,m1,lmax   
integer :: ClbIndex                  

L=2 M=1     
l1 = 1000  l2 = 1001 m1 = 578
lmax = 1024

call Clebsch2OneD(L,M,l1,l2,m1,lmax,ClbIndex)
write(*,*) ClbIndex
end program
\end{verbatim}
\titulo{Location of Example code}

\noindent \texttt{SIToolBox/examples/example\_codes/test\_Clebsch2OneD.f90}

\pagebreak{}

\subsection{CalcBipoSH}

This subroutine calculates the BipoSH coefficients from
some input map. The output BipoSH coefficients from this 
function will be in the Hazian Souradeep format, which can 
be converted to the WMAP-7 format by multiplying with 
$\frac{\sqrt{2L+1}}{\sqrt{2l+1}\sqrt{2l'+1}}\frac{1}{C_{l0l'0}^{L0}}$~\cite{Das2015}.

\titulo{Format}
\begin{verbatim}
call CalcBipoSH(Qr,Qi,LMAX,llmax,ALMll,ALMlli,Clebs) 
\end{verbatim}
\selectlanguage{english}%

\titulo{Arguments}

\begin{longtable}{>{\centering}m{0.24\columnwidth}>{\centering}m{0.12\columnwidth}>{\centering}m{0.12\columnwidth}>{\centering}m{0.4\columnwidth}}
\toprule 
Name & Kind & In/Out & Description\tabularnewline
\midrule
\endhead
\begin{verbatim}
llmax
\end{verbatim}
 & INT & IN & Maximum value that $l_{1}$ can take\tabularnewline
\begin{verbatim}
LMAX
\end{verbatim}
 & INT & IN & Maximum value of $L$ in $A_{l_{1}l_{2}}^{LM}$.\tabularnewline
\begin{verbatim}
Qr(0:dim-1),
Qi(0:dim-1)
\end{verbatim}
 & DP & IN & Real and imaginary part of $a_{lm}$s when written in a single dimensional
array.\tabularnewline
\begin{verbatim}
ALMllr(0:LMAX,0:LMAX,
0:llMAX,-LMAX:LMAX),
ALMlli(0:LMAX,0:LMAX,
0:llMAX,-LMAX:LMAX)
\end{verbatim}
 & DP & OUT & Real and Imaginary parts of the output BipoSH coefficients\tabularnewline
\begin{verbatim}
Clebs(:)
\end{verbatim}
 & DP & IN & Pre-calculated Clebsch-Gordan coefficients as a single dimensional array. \tabularnewline
\bottomrule
\end{longtable}

\titulo{Example}
\begin{verbatim}
use healpix_types    
use alm_tools    
use pix_tools    
use omp_lib

integer,parameter :: LMAX = 2    
integer,parameter :: llMAX = 1024
integer :: nside = 512
integer :: i,j,k
real(sp), allocatable, dimension(:,:) :: Map
real(dp),allocatable,dimension(:) :: Clebs
real(dp), allocatable, dimension(:) :: Qr,Qi
real(dp) :: ALMllr(0:LMAX,0:LMAX,0:llMAX,-LMAX:LMAX)
real(dp) :: ALMlli(0:LMAX,0:LMAX,0:llMAX,-LMAX:LMAX)
real(dp) :: Clebs(0:35000000)
complex(spc), allocatable, dimension(:,:,:) :: alm

...........

allocate(Qr(0:(llmax+1)*(llmax+2)/2-1))
allocate(Qi(0:(llmax+1)*(llmax+2)/2-1))
allocate(Map(0:12*nside*nside-1,1:3))
allocate(alm(1:3, 0:llmax, 0:llmax))

...........

!! READ THE MAP 

...........

dw8 = 1.0_dp    
z = (-1.d0,1.d0)
call map2alm(nside, llmax, llmax, map, alm, z, dw8) 

...........

!! READ THE CLEBSCH GORDAN COEFFICIENTS 

...........

k = 0    
do i = 0,llmax       
do j = 0,i          
Qr(k)=real(alm(1,i,j))          
Qi(k)=aimag(alm(1,i,j))          
k = k+1       
end do    
end do

call CalcBipoSH(Qr,Qi,LMAX,llmax,ALMllr,ALMlli,Clebs)

.........

end program 
\end{verbatim}

\titulo{Location of Example code}

\noindent \texttt{SIToolBox/examples/example\_codes/test\_CalcBipoSH.f90}

\pagebreak{}

\subsection{clebsch, drc3jj, drc3jm}

These are the \texttt{slatech} subroutines and these can be used for calculating the Wigner 3j symbols and the Clebsch-Gordan coefficients. 

\titulo{Format}
\begin{verbatim}
CALL DRC3JM(l, l1, l2, -m, m1min, m1max, THRCOF, NDIM, IER)

CALL DRC3JJ(l1, l2, m1, m2, l1min, l1max, THRCOF, NDIM, IER)

CALL clebsch(l, l1, l2, m, m1min, m1max, cleb, NDIM, IER)
\end{verbatim}
\selectlanguage{english}%

\titulo{Arguments}

\begin{longtable}{>{\centering}p{0.16\columnwidth}>{\centering}p{0.16\columnwidth}>{\centering}p{0.16\columnwidth}>{\centering}p{0.4\columnwidth}}
\toprule 
Name & Kind & In/Out & Description\tabularnewline
\midrule
\endhead
\texttt{l}, \texttt{l1}, \texttt{l2}
 & DP & IN & $l$, $l_{1}$, $l_{2}$ values for Wigner 3j symbol or Clebsch\textendash Gordan
coefficients\tabularnewline\\
\texttt{m}, \texttt{m1}, \texttt{m2}
 & DP & IN & $m$, $m_{1}$, $m_{2}$ values for Wigner 3j symbol or Clebsch\textendash Gordan
coefficients\tabularnewline\\
\texttt{NDIM}
 & INT & IN & Allocated dimension of the output matrix. This should be more than
or equal to the dimension of the output array of Wigner 3j symbol or
Clebsch\textendash Gordan coefficients. For example, suppose you are
calling DRC3JM for values $l=1$,$l_{1}=100$, $l_{2}=100$ and $m=1$.
So output array gives Wigner 3j for different values of $m_{1}$.
Now for this set of parameters $m_{1}$ can take $201$ different
values. So the dimension of the output array will be $201$. So NDIM
has to be more than or equal to 201. \tabularnewline\\
\texttt{m1min} , \texttt{m1max} , 
\texttt{l1min} , \texttt{l1max}
 & DP & OUT & minimum and maximum values of $m_{1}$ or $l_{1}$ that we can have
for the particular set of parameters.\tabularnewline\\
\texttt{THRCOF(1:NDIM)}, 
\texttt{cleb(1:NDIM)}
 & DP & OUT & The output array of Wigner 3j symbol or Clebsch\textendash Gordan
coefficients.\tabularnewline\\
\texttt{IER}
 & INT & OUT & Error flags :
\begin{enumerate}
\item IER=0 No errors. 
\item IER=1 Either L2.LT.ABS(M2) or L3.LT.ABS(M3). 
\item IER=2 Either L2+ABS(M2) or L3+ABS(M3) non-integer. 
\item IER=3 L1MAX-L1MIN not an integer. 
\item IER=4 L1MAX less than L1MIN. 
\item IER=5 NDIM less than L1MAX-L1MIN+1. 
\end{enumerate}

\tabularnewline
\bottomrule
\end{longtable}

\titulo{Example}
\begin{verbatim}
integer ier,i,tot
parameter (NDIM=500) 
real*8 l,l1,l2,m,m1,m2,m1min,m1max,l1min,l1max 
real*8 THRCOF(NDIM),cleb(NDIM)

l=1 l1=101 l2=100 m=1
CALL DRC3JM(l, l1, l2, -m, m1min, m1max, THRCOF, NDIM, IER)
tot = int(m1max-m1min)+1 
write(*,*)THRCOF(1:tot)

l1=101 l2=100 m1=50 m2=100
CALL DRC3JJ(l1, l2, m1, m2, l1min, l1max, THRCOF, NDIM, IER)
tot = int(l1max-l1min)+1
write(*,*)THRCOF(1:tot)

l=1 l1=101 l2=100 m=1
CALL clebsch(l, l1, l2, m, m1min, m1max, cleb, NDIM, IER)
tot = int(m1max-m1min)+1
write(*,*)cleb(1:tot)

return end
\end{verbatim}

\titulo{Location of Example code}

\noindent \texttt{SIToolBox/examples/example\_codes/test\_slatec.f90}

\newpage

\bibliographystyle{JHEP}
\bibliography{BenReference2}

\end{document}